\documentclass[twocolumn,prb,superscriptaddress,floatfix,amsmath,amssymb,showpacs]{revtex4-1}
\usepackage{graphicx}%
\usepackage{prettyref}%
	\newcommand{\pref}[1]{\prettyref{#1}}%
	\newrefformat{fig}{Fig.~\ref{#1}}%
	\newrefformat{tab}{Table~\ref{#1}}%
	\newrefformat{sec}{Sec.~\ref{#1}}%
	\newrefformat{eq}{Eq.~(\ref{#1})}%

\newcommand{\pwcond}{\texttt{PWcond}}
\newcommand{\pwscf}{\texttt{PWscf}}
\newcommand{\qe}{\textsc{Quantum~ESPRESSO}}
\newcommand{\ang}{\ensuremath{\textnormal{\AA}}}
\newcommand{\au}{\ensuremath{{\mathrm{a.u.}}}}
\newcommand{\ry}{\ensuremath{{\mathrm{Ry}}}}%
\newcommand{\ev}{\ensuremath{{\mathrm{eV}}}}%
\newcommand{\ef}{\ensuremath{E_{\mathrm{F}}}}%
\newcommand{\echem}{\ensuremath{E_{\mathrm{chem}}}}%
\newcommand{\dco}{\ensuremath{d_{\textrm{C-O}}}}%
\newcommand{\dauc}{\ensuremath{d_{\textrm{Au-C}}}}%
\newcommand{\daxc}{\ensuremath{d_{\textrm{C-axis}}}}%
\newcommand{\dwire}{\ensuremath{d_{\textrm{chain}}}}%
\newcommand{\dauau}{\ensuremath{d_{\textrm{Au-Au}}}}%
\newcommand{\pz}{\ensuremath{p_z}}%
\newcommand{\dzz}{\ensuremath{d_{3z^2-r^2}}}%
\newcommand{\dxz}{\ensuremath{d_{xz}}}%
\newcommand{\dxy}{\ensuremath{d_{xy}}}%
\newcommand{\dxx}{\ensuremath{d_{x^2-y^2}}}%
\newcommand{\sgb}{\ensuremath{5\sigma_{\textrm{b}}}}%
\newcommand{\sga}{\ensuremath{5\sigma_{\textrm{a}}}}%
\newcommand{\tps}{\ensuremath{2\pi^{\star}}}%
\newcommand{\tpa}{\ensuremath{2\pi^{\star}_{\textrm{a}}}}%
\newcommand{\tpb}{\ensuremath{2\pi^{\star}_{\textrm{b}}}}%
\begin{document}

\title{Interaction of CO with an Au monatomic chain at different strains: \
electronic structure and ballistic transport}
\date{\today}
\author{Gabriele Sclauzero}
\altaffiliation[Present address: ]{%
Ecole Polytechnique F\'ed\'erale de Lausanne (EPFL), ITP-CSEA, CH-1015 Lausanne, Switzerland}
\author{Andrea \surname{Dal Corso}}
\affiliation{%
International School for Advanced Studies (SISSA-ISAS), Via Bonomea 265, IT-34136 Trieste, Italy}
\affiliation{%
IOM-CNR Democritos, Via Bonomea 265, IT-34136 Trieste, Italy}
\author{Alexander Smogunov}
\altaffiliation[Present address: ]{%
CEA Saclay, IRAMIS/SPCSI, Bat. 462, 91191 Gif sur Yvette, France}
\affiliation{%
IOM-CNR Democritos, Via Bonomea 265, IT-34136 Trieste, Italy}
\affiliation{%
International Centre for Theoretical Physics (ICTP), Strada Costiera 11, IT-34151 Trieste, Italy}
\affiliation{%
Voronezh State University, University Sq. 1, 394006 Voronezh, Russia}

\begin{abstract}
We study the energetics, the electronic structure, and the ballistic transport of an infinite Au monatomic chain with an adsorbed CO molecule. 
We find that the bridge adsorption site is energetically favoured with respect to the atop site, both at the equilibrium Au-Au spacing of the chain and at larger spacings. 
Instead, a substitutional configuration requires a very elongated Au-Au bond, well above the rupture distance of the pristine Au chain. 
The electronic structure properties can be described by the Blyholder model, which involves the formation of bonding/antibonding pairs of $5\sigma$ and $2\pi^{\star}$ states through the hybridization between molecular levels of CO and metallic states of the chain. 
In the atop geometry, we find an almost vanishing conductance due to the $5\sigma$ antibonding states giving rise to a Fano-like destructive interference close to the Fermi energy.   
In the bridge geometry, instead, the same states are shifted to higher energies and the conductance reduction with respect to pristine Au chain is much smaller.
We also examine the effects of strain on the ballistic transport, finding opposite behaviors for the atop and bridge conductances. 
Only the bridge geometry shows a strain dependence compatible with the experimental conductance traces.
\end{abstract}

\maketitle

\section{Introduction}\label{sec:intro}

Au monatomic chains are produced in nanocontact experiments through different techniques,\cite{agrait2003} such as the Scanning Tunneling Microscope,\cite{brandbyge95,ohnishi1998} the High-Resolution Transmission Electron Microscope,\cite{rodrigues2001} and the Mechanically Controllable Break-Junction (MCBJ).\cite{yanson1998}
It has been demonstrated that a single row of suspended Au atoms between two electrical contacts typically displays a conductance of one quantum unit $G_0=2\,e^2/h$ ($e$ being the electron charge and $h$ Planck's constant),\cite{ohnishi1998} which can be satisfactorily explained by a ballistic transport model based on the Landauer-B\"uttiker theory and a single, almost perfectly transmitted conductance channel (as also confirmed by shot noise experiments, see Ref.~\onlinecite{agrait2003}). 
This conductance channel arises from the $s$-type valence electronic states of the Au chain, which form a wide, spin-degenerate electronic band crossing the Fermi level.

The reduced dimensionality of monatomic chains, and of atomic-sized contacts in general, favors the adsorption of \emph{impurities} on the atoms forming the thinnest part of the contact because of their reduced coordination with respect to bulk or surface atoms.\citep{bahn2002,valden1998,hakkinen99,Mavrikakis2000}
It is nowadays well understood that accidental or controlled contaminations with small impurities may change substantially the nanocontact conductance, because an adsorbed impurity can act as a source of electron scattering in the contact region.
On the experimental side, a large number of cases have been already examined so far, such as $\textrm{H}_2$ on Pt,\citep{smit:2002,djukic:161402,kiguchi:146802} Au,\citep{csonka2003,csonka2006} and Pd nanocontacts,\citep{csonka2004} but also on Fe, Co, and Ni nanocontacts;\citep{untiedt2004} 
CO on Pt,\citep{untiedt2004,kiguchi2007} Au, Cu, and Ni nanocontacts;\citep{kiguchi2007}
$\textrm{O}_2$ on Au and Ag nanocontacts;\citep{thijssen2006} and many others.
The effect of contamination is usually visible in the conductance histograms: the distribution of peaks often looks significantly altered after the admission of impurities in the proximity of the nanocontact.\citep{smit:2002,csonka2003,untiedt2004,kiguchi2007,kiguchi:146802,thijssen2006}
In some cases, the presence of adsorbed impurities changes the formation process and the structural properties of the chain: a remarkable example is the enhancement of monatomic chains formation for Au nanocontacts in $\textrm{O}_2$-enriched atmosphere, which has been attributed to atomic oxygen embedded in the chains.\citep{novaes2006,thijssen2006}

Several theoretical studies have been carried out in this field,\citep{Barnett2004,csonka2006,bahn2002,novaes2003,novaes2006,zhang2008,skorodumova2003,skorodumova2007,legoas2002} but the understanding of the physical properties of monatomic chains in presence of impurities is far from complete. 
In this respect, an interesting case study is the adsorption of CO on Au monatomic chains.
Recently, it has been demonstrated experimentally that CO can interact with monatomic chains produced in Au nanocontacts, modifying their conductance.\citep{kiguchi2007} 
However, compared to transition metal nanocontacts (Pt and Ni, for instance), CO only marginally alters the conductance histograms of Au nanocontacts.
CO adsorption on Au chains has also been demonstrated for short monatomic chains grown on a NiAl substrate,\cite{nilius2003} but with a much stronger effect on the electronic states of the Au chain.
Indeed, both the experimental results\cite{nilius2003} and the electronic structure calculations\cite{calzolari2004} indicate that the delocalization of some electronic states of the chain can be destroyed by a CO molecule adsorbed in the atop position.
According to ballistic conductance calculations for an unsupported chain,\cite{calzolari2004,strange2008} the impurity in the atop position acts as a ``chemical scissor'' for the coherent electron transport along the chain, resulting in a strong depression of the electron transmission near the Fermi level.
These theoretical results are clearly not adequate to satisfactorily explain the interaction of CO with Au monatomic wires and Au nanocontacts, since the conductance histogram of Au in presence of CO does not display any low conductance tail.\cite{kiguchi2007} 
The large strain imposed by the NiAl substrate on the Au chain, corresponding to a Au-Au spacing close to the rupture point of the isolated chain, and the atop position of CO might not be representative of the real adsorption geometry on suspended chains in Au nanocontacts.
Indeed, another theoretical study suggests that an adsorption geometry with a CO molecule at the bridge site of a short Au chain between surfaces gives only a very small reduction of the ballistic conductance with respect to the clean Au nanocontact.\cite{xu2006}

\begin{figure}[t]
	\begin{center}
		\includegraphics[width=0.75\columnwidth]{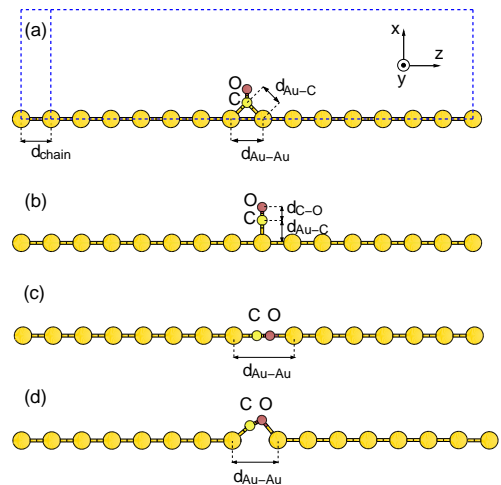}
	\end{center}
  \caption{(Color online) Sideview of the atomic geometries used to model CO adsorption on the infinite Au chain: at the (a) bridge and (b) atop sites, and in the (c) linear substitutional and (d) tilted bridge configurations.
  The tetragonal supercell is indicated for the bridge geometry with dashed lines.
  The unit cell for the metallic leads (one Au atom) is shown on the left, enclosed by dashed lines.}
	\label{fig:geom}
\end{figure}

In this work, we carry out a density functional study of the interaction between CO and a monatomic Au chain, focusing on the bonding mechanism and its effects on the electron transport properties of the chain.
The adsorption energetics of a single CO molecule at the bridge site or at the atop site are studied as a function of the Au strain.
We also consider a linear geometry with a substitutional CO, because it has been suggested as a possible intermediate meta-stable configuration responsible for the conductance increase observed while pulling a Au nanocontact contaminated with CO.\cite{kiguchi2007}
A simplified model is adopted here, consisting of an infinite, straight monatomic chain with a homogeneous spacing between the Au atoms (see \pref{fig:geom}).
We find that the bridge geometry is energetically favored with respect to the atop geometry for all Au strains considered here, while the linear substitutional configuration could occur only in presence of a very elongated Au-Au bond.

We show that the electronic structure of both the bridge and atop adsorption geometries can be interpreted through the Blyholder model\cite{blyholder1964} (and its successive refinements\cite{hammer1996,fohlisch2000}), which is commonly used to explain the interaction of CO with transition metal surfaces and has already been proven valid in the case of CO adsorbed on a Pt monatomic chain.\cite{sclauzero2008b,sclauzero2008a} 
We also find a significant correlation between the electronic transmission across the Au chain and the electronic structure, similarly to what seen for Pt chains in presence of CO.\cite{sclauzero2008b} 
For instance, a large reduction of the $s$-electron transmission is observed in correspondence of the density-of-states maximum of the $5\sigma_a$ states.
The energy and broadening of these states strongly depend on the adsorption site and on the Au strain, making the $s$ transmission rather sensible to the atomic geometry.

We also investigate how the ballistic conductance depends on the Au strain, finding two very different behaviours in the bridge and atop geometries: in the former, the conductance slightly increases with strain, while in the latter it progressively vanishes.
This is explained by the different energy of the $5\sigma_a$ states, much closer to the Fermi level (\ef) in the atop geometry than in the bridge one, and by their energy shift with strain, toward \ef\ in the atop geometry but away from \ef\ in the bridge one.
The present study is mainly aimed at a qualitative analysis of the ballistic conductance and will be successively complemented by other work to address two points which are essential to give a more quantitative picture.
First, the theoretical conductance at low Au strains suffers from a spurious contribution from $d$ electrons due to their self-interaction error, which should be tackled within a beyond-LDA approach.\cite{sclauzero2012c} 
Second, the calculated values of the tipless conductances of the bridge geometry are somewhat too large compared to the experimental values and thus call for a more realistic modeling of the nanocontact.\cite{sclauzero2012b}

The paper is organized as follows: after describing in \pref{sec:method} the numerical methods and approximations adopted in our calculations, in \pref{sec:geom} we shall present the energetics of CO adsorption on the infinite Au chain at different strains.
In \pref{sec:elstruct} and \pref{sec:ballcond} we will present the electronic structure and the strain-dependent electronic transport properties, respectively, for the tipless Au chain in presence of CO.
Finally, our conclusions will follow in \pref{sec:concl}.

\section{Methods and computational details}\label{sec:method}

We carried out first principles calculations in the framework of density functional theory\cite{hohenberg1964} (DFT) solving the single-particle Kohn-Sham (KS) equations\cite{kohn1965} for a system of atoms in periodic boundary conditions as implemented in the electronic structure code \pwscf, included in the \qe\ package.\cite{QE-2009}
We used the \citet{perdew1981} parametrization for the exchange and correlation energy within the local density approximation (LDA) and, additionally, we computed optimal distances and chemisorption energies in the generalized gradient approximation (GGA) using the Perdew-Burke-Ernzerhof (PBE) functional.\cite{perdew1996} 
Ultrasoft pseudopotentials\cite{vanderbilt1990} (USPPs) have been used to describe nuclei and core electrons.
For Au, within the LDA we use the USPP described in Ref.~\onlinecite{dalcorso2007}, while within the GGA we generated a new USPP using the reference configuration and cutoff radii of the LDA USPP.
For C and O, we use the same USPPs as in Ref.~\onlinecite{sclauzero2008b}.
The KS wavefunctions and the charge density are described using a plane wave basis set with kinetic energy cut-offs of $32 \;\ry$ and $320 \;\ry$, respectively.
Electronic occupations are broadened using a standard smearing technique,\cite{methfessel1989} with a smearing parameter $\sigma=0.01\;\ry$.
We partially relaxed our structures (see later in the text) with the Broyden–Fletcher–Goldfarb–Shanno algorithm\cite{QE-2009} optimizing selected atomic positions until atomic forces get lower than $0.026\;\ev/\ang$.

To simulate the infinite chain model with one CO molecule we use the supercell method, as exemplified in Figs.~\ref{fig:geom}(a) and \ref{fig:geom}(d) for the bridge, atop, substitutional and tilted-bridge geometries, respectively. 
Tetragonal cells with dimensions of $18 \;\au \simeq 9.5 \; \ang$ along the directions perpendicular to the wire axis and 15 Au atoms are used to compute the chemisorption energies.
The irreducible Brillouin zone (BZ) is sampled with a uniform shifted-mesh of 7 $k$-points in the direction of the wire axis and the $\Gamma$ point in the perpendicular directions.
To reduce the effects of the periodic CO replicas on the electronic bands, the number of Au atoms in the projected density of states calculations is increased to $105$ and 6 $k$-points are used to sample the BZ.

The ballistic conductance is evaluated with the Landauer-B\"uttiker formula, $G = e^2/h\: T(E_F)$, where $T(E_F)$ is the total transmission at the Fermi energy.
We calculate the electron transmission using the scattering-based approach of \citeauthor{choi1999},\cite{choi1999} generalized to the USPP scheme\cite{smogunov2004b} and implemented in the \pwcond\ code.\cite{QE-2009}
The scattering region is modeled with supercells consisting of a portion of the leads plus the impurity region, as shown in \pref{fig:geom}.
Here we use 17 Au atoms for the atop geometry and 18 atoms for the bridge geometry, but we tested longer supercells to ensure that the CO replica interaction does not affect significantly the results.
Shorter cells with one Au atom are used to compute the complex band structure of the metallic leads (leftmost portion of the supercell in \pref{fig:geom}a).

According to our LDA calculations,\cite{sclauzero2012c} and also to previous GGA calculations,\cite{delin2003} when the Au-Au spacing in the pristine monatomic chain is close to its equilibrium value, a $d$ band edge pinned at the Fermi level is responsible for a slightly magnetic ground-state and also for a large increases in the number of conductance channels, leading to a theoretical conductance of $3\,G_0$.
Both these features have never been observed in nanocontact experiments and hence can be considered a spurious effect of the LDA (or GGA).\cite{sclauzero2012c} 
A \mbox{non-magnetic} ground state and the $1\,G_0$ conductance of the Au chain can be recovered by including atomic tips in transport calculations\cite{okamoto1999,hakkinen00,delavega2004,skorodumova2005,sclauzero2012b} or considering a sufficiently large Au-Au spacing.\citep{miura2008}
It is also possible to remove these spurious features in the tipless chain by including electron correlation effects at the DFT+U level, even in a very simple formulation.\cite{sclauzero2012c}
Here we will just consider the plain LDA results for the tipless chain, focusing mainly on the qualitative features induced by CO on the transmission function, rather than on the precise ballistic conductance values.
Nevertheless, for strained geometries, where $d$ channels are ruled out from the conduction, a conductance trend with respect to strain can be extracted from our data and compared with the experimental results.

\section{Geometry and energetics}\label{sec:geom}

In order to understand if the adsorption site preference of CO may vary during different stages in the pulling process of the monatomic chain, we examine the chemisorption energy and the bonding distances at different strains of the Au chain.
Experimentally, suspended chains can be at most 7-8 atoms long and the chain edges are usually attached to the leads through tips, possibly resulting in local variations of strain owing to the different atomic coordination of the atoms at edges or in the middle of the chain.
The geometry of short suspended chains cannot be reproduce exactly with a simplified model not including the tips.
Therefore, we consider here two different limiting cases: 
in the first case, we assume that the stress is released by the elongation of the Au-Au bond where CO is adsorbed (in the bridge or in the substitutional configurations), since the interaction with CO is likely to soften that metallic bond; 
in the second case, we distribute the strain homogeneously across the chain, assuming that the Au-Au bond is not significantly weakened by the presence of the molecule.
The latter situation might be realized if the Au chain is grown on a template where the Au-Au spacing in the chain is imposed by the underlying substrate.\cite{nilius2003,calzolari2004} 
The degree of applicability of this model to the real nanocontact will be addressed in a later work.\cite{sclauzero2012b}

\begin{table}[b]
\caption{\label{tab:chainsgeom}Optimized distances \dauc\ and \dco\ (in \ang, see \pref{fig:geom}) and chemisorption energies \echem\ (in \ev) for the bridge and atop adsorption geometries.}
\begin{tabular}{c@{\hspace{13pt}}ccc@{\hspace{13pt}}ccc}
  \hline\hline
    & \multicolumn{3}{c@{\hspace{13pt}}}{Bridge} & \multicolumn{3}{c}{Atop} \\ 
 &\dauc&\dco&\echem&\dauc&\dco&\echem\\ \hline
 LDA & 1.95 & 1.16 & $-$2.4 & 1.89 & 1.14 & $-$1.0 \\
 GGA & 1.99 & 1.17 & $-$1.7 & 1.96 & 1.14 & $-$0.5\\ 
  \hline\hline
\end{tabular}
\end{table}

The CO/Au chain systems have been structurally relaxed by optimizing the positions of C and O along $x$ for the bridge and atop geometries, along $z$ for the substitutional geometry, and in the $xz$ plane for the tilted bridge geometry (cf.\ \pref{fig:geom}).\cite{sclauzero2008b}
All Au atoms are instead kept aligned along $z$ with fixed positions.
We first consider a chain with all Au-Au bonds at their equilibrium spacing, $\dwire=2.51\,\ang$ within LDA and $\dwire=2.61\,\ang$ within GGA, and we compare in \pref{tab:chainsgeom} the optimized bond lengths and the chemisorption energy (\echem) for the bridge and atop geometries.
The sum of the energies of an isolated CO and of an isolated Au chain at its equilibrium spacing is used as reference energy.\footnote{%
In the case of Au chains we have not performed fully-relativistic geometry optimizations because it has been shown in Ref.~\onlinecite{sclauzero2008b} that the SOC corrections to the chemisorption energies of CO on Pt chains are of the order of $0.1\,\ev$, or smaller, and they are not expected to be larger for Au.}
Both LDA and GGA calculations predict that the bridge site is energetically favored, by about $1.4$ and $1.2\,\ev$, respectively.
Correspondingly, the C-O bond elongation after adsorption is larger by about $0.01-0.02\,\ang$ in the bridge geometry, as was also found in the case of Pt chains.\cite{sclauzero2008b,sclauzero2008a}
As one could expected, the adsorption energies of CO on Au are generally smaller than those on Pt,\cite{sclauzero2008b} but the difference is much larger for the atop geometry.
As a result, the energy difference between the bridge and the atop site is more marked in Au chains than in Pt chains.

\begin{figure*}[tb]
    \includegraphics[width=0.85\textwidth]{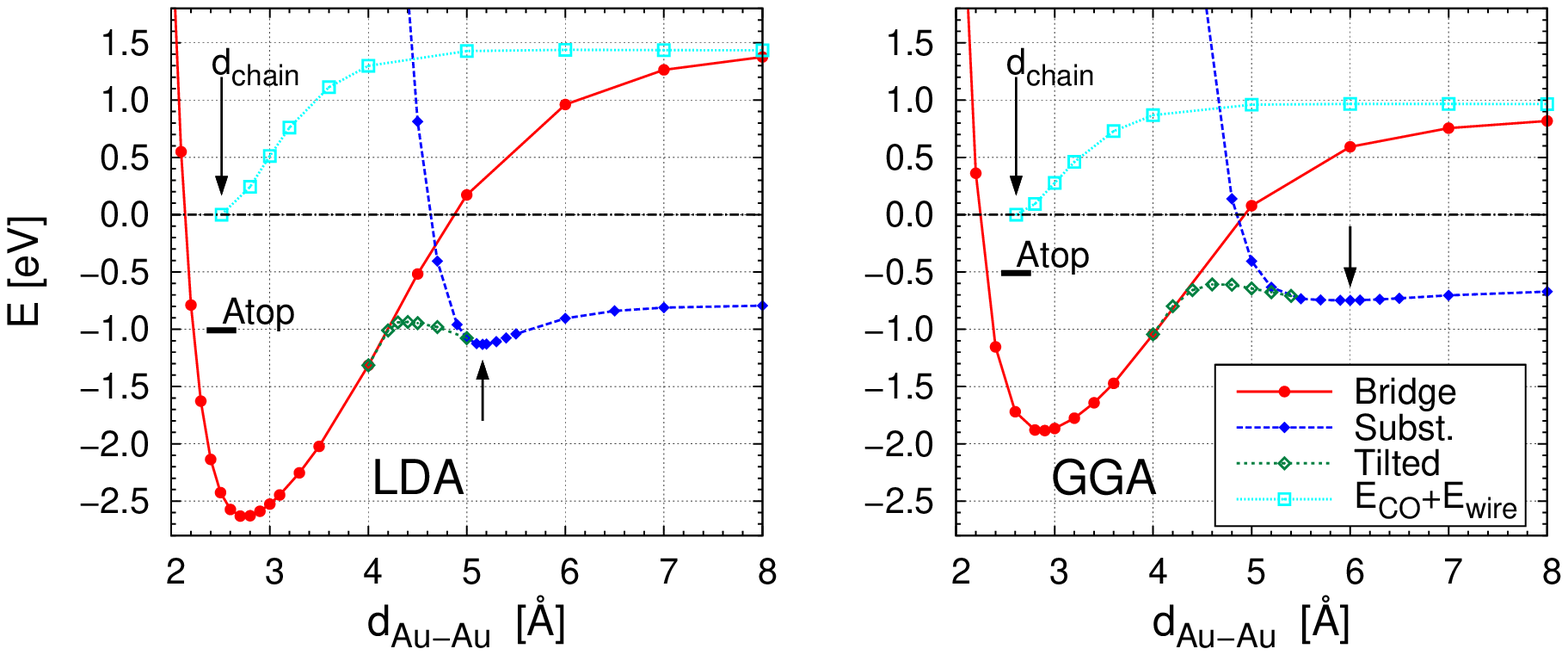}
    \caption{(Color online) Energy of the CO/Au chain system with C and O positions optimized as a function of the Au-Au distance \dauau\ (see \pref{fig:geom}).
    The bridge (filled circles), substitutional (filled diamonds), and tilted bridge (empty diamonds) geometries have been computed for several values of \dauau\ (lines through the points have been added to guide the eye), while the atop geometry (horizontal thick line) only at $\dauau=\dwire$ (see long arrow). 
The short arrows indicate the position of the energy minimum of the substitutional configuration.
The reference energy (dot-dashed line) is the sum of the energies of an isolated CO and an isolated Au chain with $\dauau=\dwire$. 
This energy has also been calculated also for some values of $\dauau > \dwire$ (empty squares).}
    \label{fig:auchainrelax}
\end{figure*}

In the first limiting case of strain, we repeat the relaxation of the atomic positions for different values of the Au-Au distance \dauau\ between the two Au atoms in contact with the molecule. 
In \pref{fig:auchainrelax}, we report as a function of \dauau\ the total energies of the bridge, substitutional, and tilted bridge geometries.
Three ranges of Au-Au bond distances can be identified: 
(i) when \dauau\ is close or slightly larger than the equilibrium spacing of the isolated chain, only the atop and bridge configurations are possible;
(ii) when \dauau\ is very large, corresponding a Au-Au bond close to or beyond the rupture point, a substitutional CO in a linear configuration is favored;\footnote{
Notice that the Au-Au spacing needed to host a substitutional CO (see also later in the text) is much larger than the theoretical breaking distance of a pristine Au chain (about $3\,\ang$, see for instance Ref.~\onlinecite{bahn2001}), hence this geometric configuration should be regarded as a possible evolution from a perpendicular CO adsorbed at the bridge or atop site.}
(iii) at intermediate values of \dauau, a tilted bridge configuration is the most stable one.

\begin{table*}[bt]
\caption{Relaxed distances (in \ang) and chemisorption energies (in \ev) of CO adsorbed at the bridge and atop sites on the infinite Au chain as obtained within LDA (GGA values in parenthesis) for selected values of the inter-atomic spacing in the chain (\dwire).}\label{tab:geomstrain}
  \begin{tabular}{c@{\hspace{13pt}}cccc@{\hspace{13pt}}ccc}
    \hline\hline
    & \multicolumn{4}{c@{\hspace{13pt}}}{Bridge} & \multicolumn{3}{c}{Atop} \\ 
\dwire& \daxc& \dauc& \dco& \echem & \dauc& \dco& \echem\\ \hline
  2.50&  1.50(1.55)&  1.95(1.99)&  1.16(1.16)&  -2.42(-1.53)&   1.89(1.96)&  1.14(1.14)&  -0.99(-0.36)\\
  2.60&  1.45(1.51)&  1.95(1.99)&  1.16(1.17)&  -2.64(-1.72)&   1.89(1.96)&  1.14(1.14)&  -1.14(-0.50)\\
  2.70&  1.41(1.47)&  1.95(2.00)&  1.16(1.17)&  -2.78(-1.83)&   1.90(1.96)&  1.14(1.14)&  -1.22(-0.58)\\
  2.90&  1.31(1.38)&  1.95(2.00)&  1.16(1.17)&  -2.91(-1.93)&   1.90(1.97)&  1.13(1.14)&  -1.26(-0.62)\\
    \hline\hline
\end{tabular}
\end{table*}

We note that the bridge geometry of CO is the lowest lying configuration not only when the Au-Au bond below the molecule is unstretched ($\dauau\simeq\dwire$), but also for a wide range of \dauau\ up to about $4.2\,\ang$ (cf.\ \pref{fig:auchainrelax}).
The energy minimum of the bridge configuration is at a distance \dauau\ larger than \dwire, as also happens in Pt chains.\cite{sclauzero2008b}
For instance, within GGA the minimum is $\dauau = 2.87\,\ang$, corresponding to an energy gain of about $0.15\,\ev$ with respect to the uniformly spaced configuration ($\dauau=\dwire$).
The regime with a very elongated Au-Au bond, corresponding to $\dauau>5.5\,\ang$ within GGA ($\dauau>5.0\,\ang$ within LDA, see \pref{fig:auchainrelax}), is characterized by an energy minimum for the substitutional configuration of CO, which is energetically favored at these very large Au-Au spacings.
Within GGA, the substitutional minimum is at $\dauau\simeq 6.0\,\ang$, a distance much larger than in the LDA ($\dauau=5.16\,\ang$), and the energy as a function of \dauau\ is much more shallow around the minimum with respect to the LDA curve.
Using as a reference the substitutional energy at $\dauau\to\infty$, the depth of the energy minimum calculated within GGA is less than $0.1\,\ev$, while in the LDA it is about $0.3\,\ev$.
Given that the substitutional minimum is much deeper in Pt than in Au (more than three times deeper according to GGA, being about $0.3\,\ev$ deep in Pt\cite{sclauzero2008b}), one might expect that, once CO is in the substitutional position, the Au-CO-Au junction could break more easily than the Pt-CO-Pt junction.
Finally, in the short range of intermediate distances corresponding to $4.3\,\ang < \dauau < 5.4\,\ang$ within GGA ($4.1\,\ang < \dauau < 5.0\,\ang$ within LDA), a tilted bridge configuration (``Tilted'' in \pref{fig:auchainrelax}) is preferred to the upright bridge and substitutional positions.
In this region, an energy maximum is found at about $\dauau=4.6\,\ang$ ($\dauau=4.4\,\ang$) between the minimum of the bridge configuration and that of the substitutional one.

To address the case where the strain is homogeneously distributed along the chain, we optimize the C and O position along $x$ for a representative set of values of the chain spacing \dwire\ between $2.5\,\ang$ and $2.9\,\ang$.
The chemisorption energies are obtained using as reference energy the total energy of the isolated Au chain at the corresponding inter-atomic spacing.
Since it has been shown that a substitutional CO can only be hosted between two very largely spaced Au atoms, here we will consider CO in the bridge and atop positions only.
In \pref{tab:geomstrain}, we report the optimized distances and the chemisoption energy \echem.
The first and second row correspond to Au-Au distances around the equilibrium lattice spacings of a pristine chain obtained from LDA ($\dwire=2.50\,\ang$) and GGA ($\dwire=2.60\,\ang$), respectively, while the third ($\dwire=2.70\,\ang$) and the fourth ($\dwire=2.90\,\ang$) describe a chain subject to a moderate tension or near rupture, respectively.
The chemisorption energies in the table show clearly that the bridge site is preferred also when the Au chain is uniformly stretched.
An increase of strain results in larger binding energies for both bridge and atop geometries, but the preference towards the bridge site is reinforced when going to larger values of \dwire.
In the bridge geometry, the increase of \echem\ with strain can be ascribed to the adjustment of the Au-C-Au angle towards the optimal value.
When the chain is strained the CO moves towards the wire axis, as seen from the decrease of the distance between the C atom and the wire axis (\daxc), while the Au-C distance (\dauc) stays almost constant.
As a consequence, the bond angle increases towards the optimal value, which corresponds to a Au-Au distance larger than the equilibrium \dwire\ (see the position of the bridge energy minimum in \pref{fig:auchainrelax}, at $\dauau \simeq 2.8\,\ang$ in LDA and at $\dauau \simeq 2.9\,\ang$ in GGA).

Since the chain is kept straight, we find a residual force acting on the Au atom below the atop CO after relaxing the C and O positions.
This force indicates a pronounced tendency for that Au atom to displace towards the C atom when the Au strain is not too high.
Indeed, by relaxing also the $x$ coordinate of that specific Au atom we find (within GGA) that it moves away from the wire axis by $0.95\,\ang$ when $\dwire=2.60\,\ang$, but only by less than $0.27\,\ang$ in a strained configuration with $\dwire=2.90\,\ang$ (in fair agreement with Ref.~\onlinecite{calzolari2004}).
At medium/high strains, \echem\ is almost stationary between $-0.65\,\ev$ and $-0.64\,\ev$ for $2.70 \leqslant\dwire\leqslant 2.90\,\ang$, and thus is only slightly larger than for the straigh chain (see GGA data in \pref{tab:geomstrain}). 
Close to the equilibrium Au-Au spacing, instead, we find $\echem=-0.72\,\ev$ at $\dwire=2.60\,\ang$, which is about $0.22\,\ev$ larger than for the straight chain.
Taking into account this additional structural relaxation, the variation of the atop \echem\ with strain is actually the opposite with respect to that of the bridge \echem.
We can argue that at lower strains the chemisorption energy is larger because the structure is more close to the optimized geometry of the neutral Au$_3$CO cluster, where the gold atoms form a nearly equilateral triangle with Au-Au distances of about $2.74\,\ang$.\cite{wu2002}

At this point it is interesting to compare our GGA data for CO adsorbed on one-dimensional chains with recent DFT calculations of CO adsorbed on Au surfaces.\cite{mehmood2009}
Although the surface calculations in Ref.~\onlinecite{mehmood2009} are at a much higher CO coverage and adopt a slightly different exchange-correlation functional (PW91 instead of PBE), the following conclusion can be drawn from the comparison:\footnote{
The theoretical nearest neighbour distance in bulk Au is around $2.94\,\ang$, so that we should compare the (211) surface data with the data for the chain at $\dwire=2.90\,\ang$.}
(i) distances and chemisorption energies for CO atop of step edge atoms in the Au(211) surface are not much different from those in the monatomic chain (surface: $\dauc=1.99\,\ang$, $\echem=-0.54\,\ev$; chain: $\dauc=1.97\,\ang$, $\echem=-0.62\,\ev$); 
(ii) for CO adsorbed at the bridge site the differences are larger, especially for \echem\ (surface: $d_{\textrm{C-surf}}=1.46\,\ang$, $\echem=-0.65$; chain: $d_{\textrm{C-axis}}=1.38\,\ang$, $\echem=-1.93$)
(iii) larger chemisorption energies in the chain are attributable to different coordination numbers of Au atoms, much smaller in the chain, since the comparative study of the different Au surfaces shows that a decrease in the local coordination of surface atoms leads to higher binding energies of CO;\cite{Mavrikakis2000,bahn2001,mehmood2009}
(iv) since the \echem\ difference with respect to the surface case is much larger for CO at the bridge than for CO atop, the preference for the bridge site in the monatomic chain is much more marked than in the stepped surface (where the two sites are separated by just $0.11\,\ev$); 
(v) the C-O bond elongation in the surface case ($\dco=1.17\,\ang$ for CO atop and $\dco=1.15\,\ang$ for CO at the bridge) is comparable to the values obtained for the one-dimensional chain.

\section{Electronic structure}\label{sec:elstruct}

In this section we will analyze the projected density of states (PDOS) of the bridge and atop geometries at the Au equilibrium spacing (\pref{tab:chainsgeom}).
We consider the projections onto the atomic orbitals of C, O and of the Au atom(s) in contact with the molecule.
We report only the projections onto atomic orbitals which are even with respect to the mirror symmetry through the plane containing both the CO and the chain.\cite{sclauzero2008b} 
Indeed, it was already shown for Pt chains that the even PDOS includes the interaction of the metal with both the $\sigma$ and even $\pi$ molecular levels of CO, while the odd PDOS involves only the odd $\pi$ interaction.\cite{sclauzero2008b}
There are actually slight differences between the even and the odd $\pi$ interactions, which however do not change qualitatively the conclusions obtained by examining the even PDOS only.\cite{sclauzero2010}

\begin{figure}[tb]
    \includegraphics[width=0.75\columnwidth]{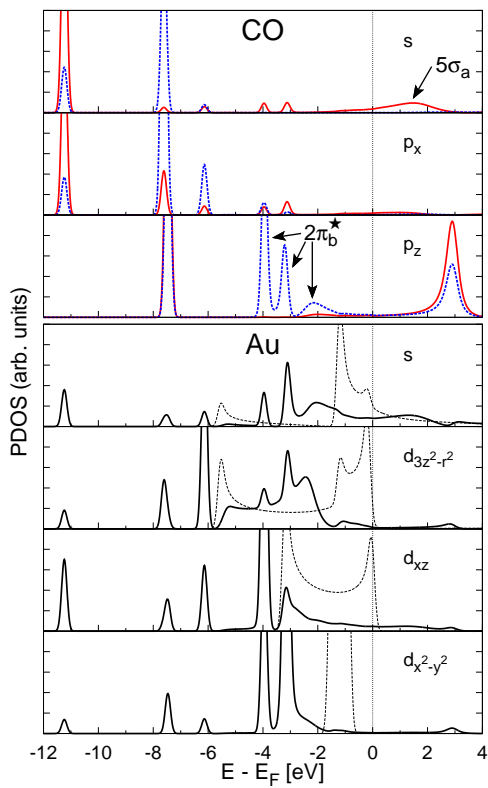}
    \caption{(Color online) PDOS for the LDA bridge configuration.
   In the upper panel, we project onto the even atomic orbitals of C (solid red lines) and O (dashed blue lines).
   In the lower panels, the projections are on the even orbitals of one of the two Au atoms close to CO (solid lines). The PDOS for the isolated Au chain are also shown (dashed lines).}
    \label{fig:pdosSRbridgeAu}
\end{figure}

\subsection{CO at the bridge site}
The even PDOS for the bridge geometry are shown in \pref{fig:pdosSRbridgeAu}, the projections on the even atomic orbitals of C and O ($s$, $p_x$, and $p_z$) being in the upper part of the figure and those on the even atomic orbitals of one of the two (equivalent) Au atoms below CO ($s$, \dzz, \dxz, and \dxx) in the lower part.
In the PDOS projected onto the $s$ and $p_x$ orbitals of C and O we can distinguish the peaks corresponding to the original $\sigma$ orbitals of CO. 
The $3\sigma$ level is very low in energy (at $-23.5\,\ev$, not shown) and does not take part in the interaction, while the $4\sigma$ and $5\sigma$ levels at $-11.2\,\ev$ and $-7.7\,\ev$, respectively, have a moderate hybridization with Au states, as witnessed by the small peaks at the same energies in the Au PDOS.
The interaction between CO and the Au chain produces new $\sigma$-derives states, absent in the isolated molecule, similarly to what was observed in a previous study of CO on Pt chains.\cite{sclauzero2008b}
In the $s$ PDOS of C there are three small additional peaks below \ef\ and also a very broad feature, centered at about $1.5\,\ev$ above \ef, which we will call \sga.
The latter corresponds to a $\sigma$-charge donation from the molecule to the metal.\cite{blyholder1964,hammer1996,fohlisch2000}
The \sga\ state is mainly coupled with the metal $s$ states, hence we might expected that it will strongly perturb the $s$ channel transmission at the corresponding energies, analogously to what observed for Pt chains.\cite{sclauzero2008b}

Among the $\pi$-derived states, visible in the $p_z$ PDOS of C and O, the $1\pi$ and \tps\ molecular states, centered at $-7.4\,\ev$ below \ef\ and at $2.9\,\ev$ above \ef, respectively, have a small hybridization with Au states and correspond to small peaks in the metal PDOS.
In the middle of the $1\pi$ and \tpa\ peaks, we find three separate peaks between $-4\,\ev$ and $-2\,\ev$ with distinct heights and widths.
These states, which we call \tpb\ states, arise from the backdonation of Au electrons which partially fill the formerly empty \tps\ levels of the molecule.\cite{blyholder1964,hammer1996,fohlisch2000} 
The three separate contributions to the \tpb\ DOS are due to hybridizations of CO $\pi$ states with different even orbitals of Au, as visible from the relative intensities of the peaks in the Au PDOS.
At larger binding energies, the two narrow peaks at about $-4.0\,\ev$ and $-3.2\,\ev$ can be associated to states having mostly \dxz\ character or mostly \dxx\ character, respectively, while the bump at smaller energies (about $-2.1\,\ev$) corresponds to states having predominant $s$ character. 
Thus, a donation/backdonation process takes place between CO and the Au chain, similarly to what found for Pt chains.\cite{sclauzero2008b}
In both Pt and Au chains, this process leads to the formation of bonding/antibonding pairs, \sgb/\sga\ and \tpb/\tpa, with the \sga\ lying mainly above \ef\ and the \tpb\ below \ef.

\begin{figure}[tb]
    \includegraphics[width=0.75\columnwidth]{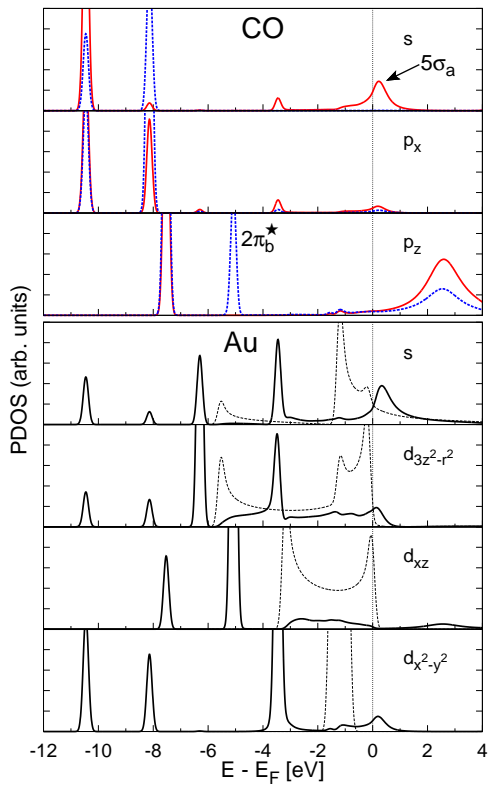}
    \caption{PDOS for the LDA atop configuration. 
    The plots are organized in the same way as in \pref{fig:pdosSRbridgeAu} and show the PDOS of C and O (upper panels) and of the Au atom below CO (lower panels).}
    \label{fig:pdosSRontopAu}
\end{figure}

\subsection{CO at the atop site}
In \pref{fig:pdosSRontopAu}, we report the even PDOS for the atop geometry, including the projections onto the atomic orbitals of C, O, and of the Au atom below the CO.
As seen in the PDOS for the brigde geometry, in the $s$ and $p_x$ PDOS of C and O we find the $4\sigma$ and \sgb\ peaks below \ef, as well as the additional \sga\ peak arising from the charge donation process above \ef.
The \sga\ peak still matches with a peak in the $s$ PDOS of the metal as in the bridge geometry, but it is less broadened by the interaction and it is found much closer to \ef, at about $0.2\,\ev$.
We have already shown for Pt that the \sga\ peak is correlated to a dip in the $s$-electrons transmission.\cite{sclauzero2008a,sclauzero2008b} 
This dip did not influence the ballistic conductance of the Pt chain because it was located at $1.2\,\ev$ and at $2.1\,\ev$ above \ef\ in the atop\cite{sclauzero2008a} and in the bridge geometry,\cite{sclauzero2008b} respectively, and also because of the large contribution of $d$ electrons to the conductance.
At variance with Pt, only $s$-type electrons contribute to the Au conductance,\cite{sclauzero2012c} therefore one can expect that in the atop configuration the \sga\ transmission dip will also modify the ballistic conductance of the Au chain.
Finally, we notice an additional peak at $-3.5\,\ev$ which strongly perturbs the $s$, \dzz, and \dxx\ PDOS of Au around those energies, but has a much smaller intensity in the $s$ and $p_x$ PDOS of C and O.

Among $\pi$-derived even states, the distribution of peaks in the $p_z$ PDOS of C and O is characterized 
by three peaks: the narrow $1\pi$ and \tpb\ peaks below \ef\ at $-7.6\,\ev$ and at $-5.1\,\ev$, respectively, and the broader \tpa\ feature above \ef, centered at about $2.6\,\ev$.
A single even \tpb\ peak was also found at similar binding energies in the PDOS of the Pt chain with CO in the same position.\cite{sclauzero2008a,sclauzero2010} 
This consideration suggests that some details of the even $\pi$ interaction depend mainly on the adsorption geometry, rather than on the metal.
Indeed, in the atop geometry even $\pi$ states are coupled prevalently with the \dxz\ orbitals of the metal atom, essentially for geometrical reasons and orbital overlaps, resulting in a single peak.
In the bridge geometry instead, there is a strong hybridization also with the other metal orbitals, giving rise to three distinct contributions (which in Pt merge together to give a broad peak), each characterized by a different strength of the coupling to \dxz, \dxx, and $s$ orbitals.

As seen for Pt chains, the onset of these new localized states brought by the interaction with CO perturbs strongly the overall density distribution in the PDOS of the metal atoms in contact with the molecule, especially for states associated to narrow electronic bands, such as the \dxx\ and the \dxy.
The width of $|m|=1$ and $|m|=2$ bands in Au chains is visibly reduced with respect to Pt chains and the depletion of density due to CO adsorption is consequently enhanced in Au.
Notably, the $|m|=2$ bandwidth in Au is reduced by about a half with respect to Pt and the \dxx\ and \dxy\ PDOS suffer a much larger reduction of the density at energies between $-1.2\,\ev$ and $-0.1\,\ev$.
Therefore, at electron scattering energies between $-3.3\,\ev$ and \ef\ we could expect a greater reduction in the transmission of Au chains (examined in the next section) with respect to that of Pt chains.\cite{sclauzero2008a,sclauzero2008b}

\section{Ballistic transmission}\label{sec:ballcond}

In this section, we report the LDA electron transmission of the bridge and atop geometries and discuss the effects of the uniform Au strain. 
As explained in \pref{sec:method}, at low strains the tipless ballistic conductance results are affected by the spurious contribution from the $d$ states at the Fermi level, hence only a qualitative discussion of the transmission will be presented for Au-Au spacings close to the equilibrium value. 
The conductance trend with respect to strain can be anyway inferred from these tipless transmissions because at larger strains the $d$ channels move to lower energies and do not contribute to the conduction. 

\begin{figure*}[t]
  \includegraphics[width=1.0\textwidth]{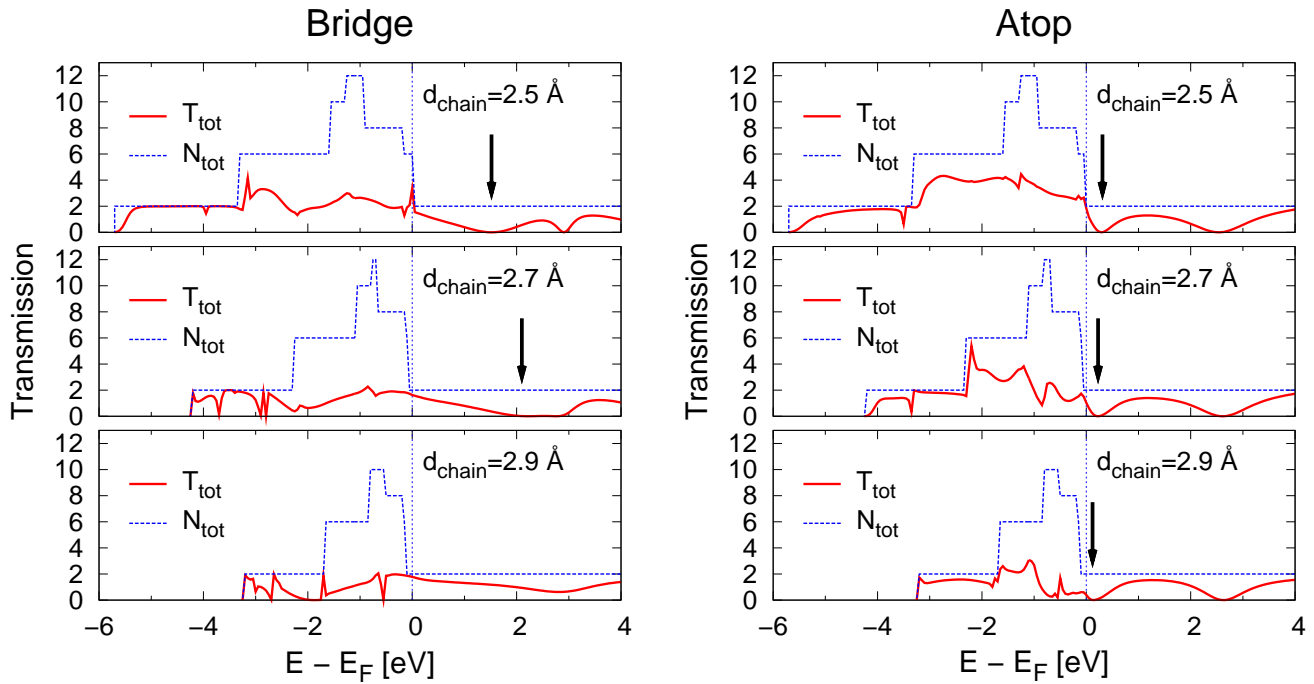}
  \caption{(Color online) Total LDA transmission (solid red lines) and number of channels (dashed blues lines) for a tipless Au monatomic chain with CO adsorbed at the bridge or at the atop site (left and right side, respectively).
  Different inter-atomic Au spacings (\dwire) are considered and the corresponding optimized distances given in \pref{tab:geomstrain} are used.
  The arrows point the position of the \sga\ transmission dip (see text).}\label{fig:condstrain}
\end{figure*}

In \pref{fig:condstrain}, we report the transmission (solid lines) and the number of available channels (dashed lines) as a function of the electron scattering energy for the bridge and atop geometries.
We first consider the unstrained chain with $\dwire=2.5\,\ang$ (panels at the top of the figure) and make a connection between the electron transmission and the electronic structure presented in the previous section.
Starting from the lowest scattering energies, we see that only one spin-degenerate $m=0$ channel is available below $-3.3\,\ev$.
This channel has mostly \dzz\ character and is well transmitted in both the bridge and the atop geometry, except for the smoothing at the bottom of the $m=0$ band and a narrow dip at higher energies. 
This transmission dip matches with a specific peak in the $s$, \dzz, and \dxx\ PDOS of Au, at $-3.5\,\ev$ in the atop geometry (\pref{fig:pdosSRontopAu}) and at $-4.0\,\ev$ in the bridge geometry, in correspondence of the lower \tpb\ peak (\pref{fig:pdosSRbridgeAu}).

For energies below $-3.3\,\ev$, the transmission is slightly larger when CO is at the bridge and in fact, excepting the lower band edge region, the \dzz\ PDOS in the bridge geometry is less perturbed than in the atop geometry.
Conversely, at higher energies, the bridge transmission is lower than the atop transmission in the whole energy range of the $|m|=1$ band ($-3.3< E < 0.0\,\ev$), especially for energies around $-2.2\,\ev$.
In the bridge geometry, the wide depression of the transmission centered at that energy can be associated with the broader \tpb\ peak in the \pz\ PDOS.
This peak is more coupled to $s$ and \dzz\ states and is absent when CO is atop because of the different orbital overlaps discussed in \pref{sec:elstruct}.
The $|m|=2$ scattering states are available only between $-1.7\,\ev$ and $-0.9\,\ev$ and are completely blocked in presence of CO for both geometries.
The $|m|=2$ PDOS of the Au atom(s) in contact with the molecule are actually strongly perturbed and completely detuned from the density of the narrow $|m|=2$ band of the chain (see, for instance, the \dxx\ PDOS of Au in \pref{fig:pdosSRbridgeAu}).

Above the Fermi level, for energies up to $4\,\ev$ only one spin-degenerate $m=0$ channel (mostly of $s$-type) is available for transmission.
In this energy region the electron transmission is characterized by the presence of two wide dips. 
They are present in both bridge and atop geometries but are centered at different energies which match exactly with the positions of \sga\ and \tpa\ peaks in the corresponding PDOS. 
These two Fano-like features in the transmission can be interpreted as a consequence of a destructive interference between two electron propagation paths: 
one going directly along the Au chain and another one passing through the corresponding resonance (\sga\ or \tpa) brought by the binding of CO to the Au chain.
We notice that in the atop geometry the \sga\ transmission dip, pointed by the arrows in \pref{fig:condstrain}, falls just $0.2\,\ev$ above the Fermi level, while in the bridge geometry it is located at about $1.6\,\ev$.
The different position of the \sga\ dip turns out to be critical for the ballistic conductance when considering a strained chain geometry (see later) or when correcting for the self-interaction error of the $d$ states,\cite{sclauzero2012c} since the position of the $d$ band edge responsible for the additional (spurious) contributions to the conductance is very sensitive to these factors.

The effects of strain on the electron transmission are described in the central and bottom panels of \pref{fig:condstrain}, with the strain increasing when going from the central to the bottom panels.
The number of available channels at a given scattering energy (dashed lines) can vary with strain, reflecting the changes in the electronic band structure of the isolated infinite Au chain upon stretching:
(i) the bandwidths are progressively reduced as the Au-Au spacing increases; and
(ii) the center of each band moves in energy with respect to the Fermi level.
As a consequence, the positions of the $d$ bands edges are shifted with respect to \ef\ and so are the ranges of availability of the corresponding transmission channels. 
In particular, the number of open conductance channels changes as a function of the bond length:\cite{delin2003} at moderate strains ($\dwire=2.70\,\ang$, see central panels of the figure), the $|m|=1$ $d$-channels have already moved below \ef, leaving only one spin-degenerate $s$-channel available for conductance.
However, a further increase of the strain will not change the number of conductance channels anymore.

Also the transmission in presence of CO undergoes important modifications when the chain is stretched (solid lines in \pref{fig:condstrain}).
In the bridge geometry, a moderate strain introduces many dips in the transmission of $m=0$ states below $-2.3\,\ev$ and enhances the depression due to \tpb\ states, which is still centered at about $-2.1\,\ev$.
Around the Fermi energy, the $d$ band edge resonance is removed and the transmission looks smoother.
Above \ef, the \sga\ dip is shifted higher in energy, following the shift of the corresponding peak in the PDOS (not shown here), and merges with the \tpa\ dip to form a wide region where the transmission is strongly suppressed. 
At higher strains ($\dwire=2.90\,\ang$) the transmission at energies below $-0.6\,\ev$ degrades further, while above that energy it actually increases so that the ballistic conductance slightly increases too, from $0.81\,G_0$ to $0.88\,G_0$.
Interestingly, the merging of the \sga\ and \tpa\ dips in the transmission leads to the partial restoring of the ideal transmission of the Au chain around $2.8\,\ev$, which can be thought of as a result of mutual cancellation of two destructive interferences.
In the atop geometry (right hand side), an increase of strain produces a progressively stronger transmission reduction close to the Fermi level and also for lower energies, down to $-2\,\ev$. 
At variance with the bridge geometry, above \ef\ the presence of the \sga\ and \tpa\ dips persists separately also at high strains.
In particular, the \sga\ dip gradually approaches \ef, causing a progressive reduction of the conductance with increasing strain.
Therefore, the bridge and atop geometries display an opposite behaviour of the conductance with strain.

\section{Conclusions}\label{sec:concl}

In summary, we have investigated the adsorption of CO on a Au monatomic chain within density functional theory and we have studied how the electronic structure and the ballistic transport properties of the Au chain are modified by the interaction with CO.
Using an infinite chain geometry we have described the energetics of CO in a few possible adsorption sites (bridge, atop, and substitutional) and, for two of them (bridge and atop), we have also studied its dependence on the strain of the Au chain.
We find that the upright bridge position of CO is energetically favored, while a linear substitutional configuration becomes convenient only in presence of a very elongated Au-Au bond, well beyond the break distance for the pristine monatomic chain.
Thus, the site preference ordering for CO on Au chains is the same as for Pt chains,\cite{sclauzero2008a,sclauzero2008b} but chemisorption energies in Pt are substantially larger than in Au (at least $0.6\,\ev$).
The chemisorption energies increase with the Au strain and the site preference prediction towards the bridge site is reinforced at higher strains.

The electronic structure of the system can be explained by the Blyholder model,\cite{blyholder1964,hammer1996,fohlisch2000} in terms of the donation and backdonation of electrons between the molecule and the metal.
The adsorption process produces a broadening and a strong hybridization of the highest occupied and lowest unoccupied levels of the molecule, resulting in the formation of bonding-antibonding pairs of $5\sigma$ and \tps\ states: the presence of antibonding \sga\ (bonding \tpb) states above \ef\ (below \ef) can be associated to the electron donation (backdonation). 
These new states originated from the CO-metal interaction appear as distinct peaks in the density of states of the bridge and atop geometries and their binding energies, intensity, broadening, as well as the details of the coupling to the metal states depend strongly on the adsorption site of CO.

In general, the electronic structure influences the transmission across the chain at all scattering energies interested by the interaction with CO, as already seen for Pt chains.\cite{sclauzero2008a,sclauzero2008b}
Among Au $d$ states, those showing a stronger coupling with the CO levels are more poorly transmitted than the others, while $s$ states are generally well transmitted except for some particular energies in correspondence of the hybridization peaks in the PDOS. 
Some dips in the transmission can be easily associated to peaks appearing in the PDOS at the same energies, similarly to the observed correspondence between the dips in the transmission across a molecule with a side chain and the molecular orbital energies of the isolated side chain.\cite{ernzerhof2005}
In particular, the \sga\ peak can be associated to a dip or a depression in the $s$-electron transmission.
In the atop geometry, the \sga\ peak falls very close to the Fermi level (\ef), while in the bridge geometry it is found well above \ef, hence the conductance reduction due to CO displays a marked dependency on the adsorption site.
For Au-Au spacings well above the equilibrium value of the pristine chain, the conductance is considerably smaller in the atop geometry because the $s$ channels are the only available for electron transport at \ef.
At Au-Au spacings close to the equilibrium, instead, the conductance values from plain LDA calculations are subject to a large error due to the wrong position of $d$ states\cite{sclauzero2012c} and do not allow a clear comparison between the two geometries.

We have also studied how the transport properties of the chain in presence of CO depend on the strain.
The transmission of $d$ channels degrades at increasing strains, while that of the $s$ channel is characterized by the presence of dips which can appear, disappear or move in energy, depending on the adsorption site.
Excluding the low strain conductance value affected by the spurious $d$ channels,\cite{sclauzero2012c} the bridge and atop conductances behave differently as the chain gets stretched: the former slightly increases with strain, while the latter decreases because of the \sga\ dip approaching to \ef.
Thus, the bridge geometry, besides being energetically favoured, has a strain dependence of the conductance compatible with the conductance traces of Au nanocontacts in presence of CO, which show a conductance increase while the Au tips are pulled apart.\cite{kiguchi2007}
The conductance of the substitutional geometry at the equilibrium Au spacing is found to be much smaller than that of the bridge geometry,\cite{sclauzero2010} and therefore it is unlikely that a tilting of the molecule from the bridge to the substitutional position could result in a conductance increase upon contact stretching.

\begin{acknowledgments}
The authors are grateful to Erio Tosatti for useful and stimulating discussions.
This work has been supported by PRIN-COFIN 20087NX9Y7, as well as by INFM/CNR ``Iniziativa trasversale calcolo parallelo''. 
All calculations have been performed on the SISSA-Linux cluster and at CINECA in Bologna.
\end{acknowledgments}


%

\end{document}